\documentclass[runningheads,a4paper]{llncs}

\usepackage{amssymb}
\usepackage{graphicx}
\usepackage{float}
\usepackage{caption}

\usepackage{url}
\urldef{\mailsa}\path|{alfred.hofmann, ursula.barth, ingrid.haas, frank.holzwarth,|
\urldef{\mailsb}\path|anna.kramer, leonie.kunz, christine.reiss, nicole.sator,|
\urldef{\mailsc}\path|erika.siebert-cole, peter.strasser, lncs}@springer.com|    
\newcommand{\keywords}[1]{\par\addvspace\baselineskip
\noindent\keywordname\enspace\ignorespaces#1}

\usepackage{color, colortbl}
\definecolor{Gray}{gray}{0.8}

\begin{document}

\mainmatter  

\title{OpenTED Browser: \\Insights into European Public Spendings}


%
%
\author{Yann-A\"el Le Borgne\inst{1}
\and Adriana Homolova\inst{2} 
\and Gianluca Bontempi\inst{1}}
%

\institute{Machine Learning Group, Universit\'e Libre de Bruxelles\\
Boulevard du Triomphe, CP212\\ 1050 Brussels, Belgium\\
\and
Independent Data Journalist, Amsterdam, Netherlands\\
}

%
%

\maketitle

\begin{abstract}
We present the \emph{OpenTED browser}, a Web application allowing to interactively browse public spending data related to public procurements in the European Union. The application relies on Open Data recently published by the European Commission and the Publications Office of the European Union, from which we imported a curated dataset of 4.2 million contract award notices spanning the period 2006-2015. The application is designed to easily filter notices and visualise relationships between public contracting authorities and private contractors. The simple design allows for example to quickly find information about who the biggest suppliers of local governments are, and the nature of the contracted goods and services. We believe the tool, which we make Open Source, is a valuable source of information for journalists, NGOs, analysts and citizens for getting information on public procurement data, from large scale trends to local municipal developments.

\end{abstract}

\section{Introduction}

Public procurement is the process whereby governments buy goods and services, such as office supplies, equipment, buildings, roads, and so forth. It represents about one third of total government expenditures in OECD countries \cite{cernat2015international}. This is usually a public-private deal in which the buyer is a public entity and the (winning) bidder is usually a privately owned company. Public procurements are mostly financed by public funds \cite{digiwhistD11,hoekman2015international}. 

The development of Tenders Electronic Daily (TED) \cite{TED} can be considered as the biggest EU-wide effort made so far to support procurement across borders. TED, which is managed by the Publications Office of the European Union \cite{PO}, is the online version of the S Series of the Official Journal of the European Union (OJEU), which is a supplement to the Journal particularly focused on European public procurement. TED publishes over 1000 new over EU-threshold value procurement notices every day worth about EUR 400 billions a year  \cite{cernat2015international,de2014alternative,DGGROWG4}. Furthermore, TED also publishes other documents coming from funds to be spent on EU external aid and the procurement of EU institutions.

Along with TED, the Publications Office provides bulk downloads of its data. The raw dataset contains all contract notices for tender data since 1995 in XML format, and its size is around 100 GB (GigaBytes). The browsing and analysis of this dataset raises a number of complex challenges, due to varying quality of data between countries and years, missing values, variability in the naming of contracting authorities and entities, multilingual documents, and so forth. 

The Directorate-General for Internal Market, Industry, Entrepreneurship and SMEs (DG-GROW) of the European Commission has recently initiated great efforts in the curation of the Publications Office data, and released in August 2015 a summary of all contract award notices (CANs) for European public procurements for the period spanning 2006-2015 \cite{TEDDataSet}. The curated dataset provides an unprecedented source of information on public spending in the European Union and yields very valuable information on procurement data. In particular, it provides information related to the identity of contracting authorities and contractors, the nature of the supplies/works/services, the final contract values and the number of offers for a given contract notice. Such information not only makes it possible to provide insights in the network of public/private partnerships, but also to exhibit procurements patterns across all European countries, or to detect (and avoid) corruption \cite{alvarez2012towards,de2014alternative,miroslav2014semantic,uyarra2013review}.

The relative large size of the dataset however still prevents its analysis from users without an analytics background: Data is provided as CSV files for each year of CAN (from 2006 to 2015, ten files in total), whose size ranges from around 100 MB (MegaBytes) to 300 MB, totalising about 2 GB (GigaBytes). While the size of the dataset can be stored without trouble on current laptop or desktop configurations, the opening of such files remains a challenge for standard spreadsheet applications such as Excel, and makes analyses spanning multiple years (i.e. filtering data from multiple files) very tedious. It must be stressed that most of the people interested in CAN data (journalists, entrepreneurs, citizen, ...) do not have the necessary analytics background to explore datasets of such size.   

The OpenTED browser aims at bridging this gap, and our contributions are the following. First, data is stored on our OpenTED server, and does not need to be fully downloaded, making it easier for users with slow Internet connection to access the data. Users can furthermore filter the data they need (according to countries, years, type of goods, and so forth), and download only what they are interested in. Second, we provide a visualisation tool, based on Sankey diagrams, that represents as a graph the contract awards between public authorities and private contractors. The visualisation makes clear how much money the contracts are worth, and provides hyperlinks to the official TED award notices for further details. Finally, we make the code for both the data preprocessing (Python) and the Web application (R Shiny) open source. A docker container can be used to run the Web application on a local machine, making the interaction with the application even faster.

The paper is organised as follows. Section 2 details the official TED CAN data, and how they were preprocessed for the OpenTED browser. Section 3 presents the OpenTED browser Web interface. Section 4 presents the lottery game `Public spending is fun! who wants to be a supplier?', a `gamification' of the search for contract award notices. Section 5 concludes the paper with open issues and perspectives. 

\section{Data and Methods}

Raw data was retrieved from the Open Data portal of the European Union \cite{TEDDataSet}, slightly preprocessed for easier querying, and converted to Parquet format for efficiency reasons. We describe the data and preprocessing hereafter.

\subsection{Data overview}

The data comes from the European Economic Area, Switzerland, and the former Yugoslav Republic of Macedonia  and covers the time period between 2006/01/01 and 2015/12/31. It is in comma separated value (CSV) format and is encoded as UTF-8. Generally, the data consists of notices above the procurement thresholds. However, publishing below threshold notes in TED is considered good practice, and thus a non-negligible number of below threshold notices is present as well  \cite{TEDDataSet}. 

\begin{figure}
\centering
\includegraphics[width=1\textwidth]{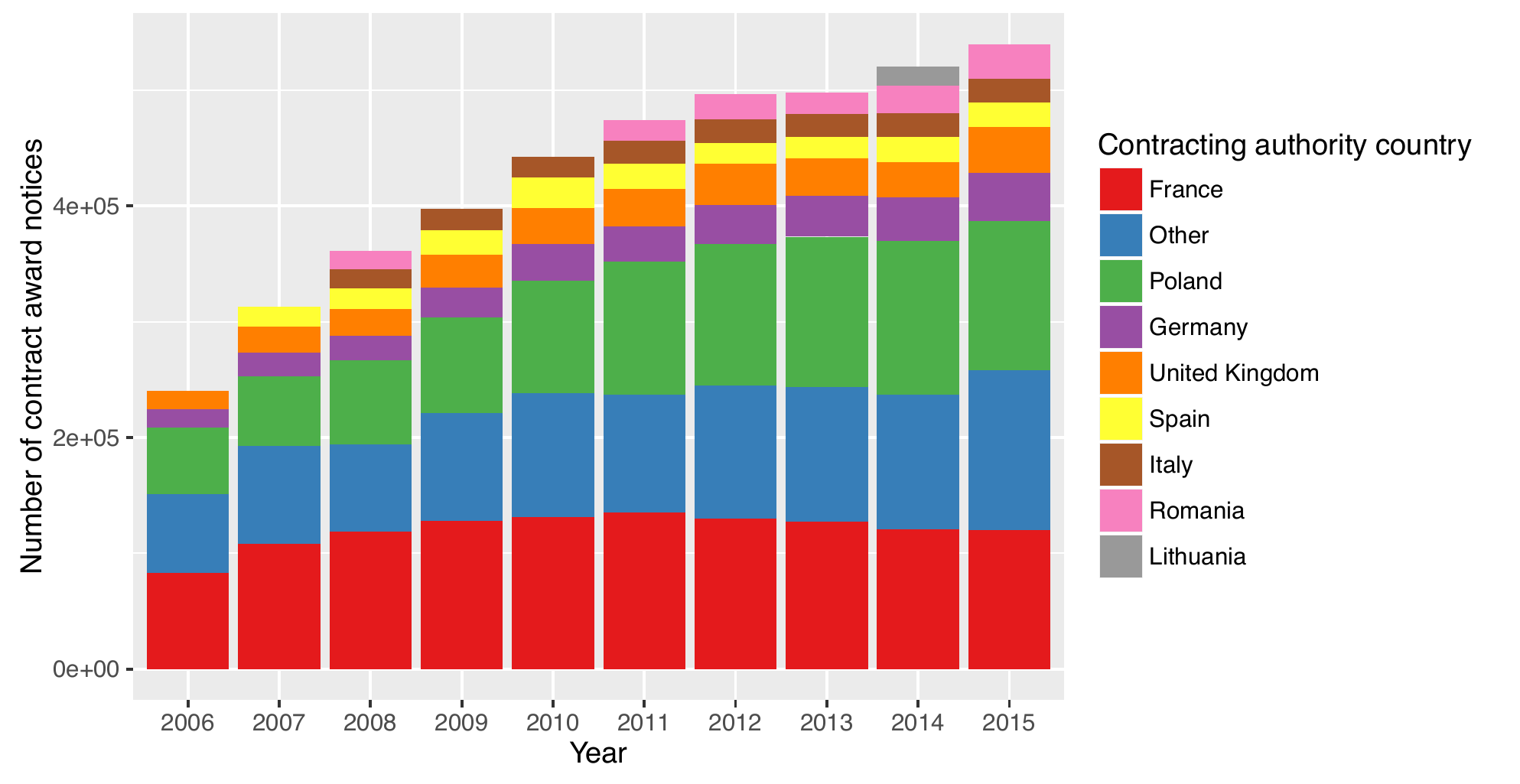}
\caption{Number of contract award notices (CANs) per country and per year, since 2006. Only countries with more than 15000 CANs per year are reported (number of CANs for other countries are summarized in `Other'.}
\label{CANperCountryPerYear}
\end{figure}

The number of countries covered has increased throughout the years, generally in line with their accession to the single European market. This is illustrated in Fig. \ref{CANperCountryPerYear}, which reports  the number of CANs per country and per year, since 2006. Only countries with more than 15,000 CANs per year are reported for clarity reasons (number of CANs for other countries are summarized in `Other').

The number of CANs increased from about 250,000 in 2006 to more than 500,000 in 2015, and a total of 4,283,986 CANs are available in the dataset. It is worth noting that the number of CANs available per country is quite imbalanced, and are the highest for France, Poland, Germany and the United Kingdom (UK).

The data includes 48 selected fields from CANs (Table \ref{CANdata} in the Annex), divided in six categories: \emph{Notice metadata}, \emph{Contracting authority or entity identification}, ￼￼\emph{Winning bidder identification}, \emph{Various CAN level variables}, and \emph{Various CA level variables} \cite{TEDDataSet}. The size of the CSV file with all notices is slightly above 2GB.

\subsection{Data preparation and conversion to Parquet}

We remained as faithful as possible to the original data, and included all 4,283,986 records and 48 fields in the OpenTED browser. For making the filtering of the dataset more user-friendly, we however renamed nine fields, which are highlighted in the OpenTED browser (see Section \ref{TEDBrowser}). These fields consist in the date of the CAN, the identifier (ID), the name and country of the contracting authority, the name and country of the contractor, the value of the CAN, the number of offers and the CPV code. The renaming is detailed in Table \ref{CANdata2}.

\begin{table}[ht]
\centering
\caption{Renaming of highlighted fields in the OpenTED browser.}
      \scriptsize
      \begin{tabular}{|l|l|l|}
        \hline
         Original name   & Description & New name\\ \hline
        \textbf{DT\_DISPATCH} & The date when the buyer  & Dispatch\_Date\\
        &dispatched (sent) the notice&\\
        \textbf{ID\_NOTICE\_CAN} & Unique identifier of the contract award  & Award\_Notice\_Id\_Link\\
        &notice for publication to TED & \\ 
        \textbf{CAE\_NAME} & Official name & Contracting\_Authority\_Name \\ 
       \textbf{ISO\_COUNTRY\_CODE} & Country &  Contracting\_Authority\_Country \\
         \textbf{WIN\_NAME} & Official name & Contractor\_Name  \\
          \textbf{WIN\_COUNTRY\_CODE}& Country& Contractor\_Country \\
          \textbf{VALUE\_EURO} & CAN value, in EUR, without VAT. & Contract\_Value\_Euros\\
& If the value was not present &\\
&, the lowest bid is included&\\
        \textbf{NUMBER\_OFFERS}& Number of offers received & Number\_Offers\_Received\\
          \textbf{CPV} &  The main Common Procurement  & CPV\_Code\\
&Vocabulary code of the main &\\
&object of the contract &\\
        \hline
      \end{tabular}
\label{CANdata2}
\end{table}

Besides the renaming, we also reformatted fields involving dates, countries, and values. In the original data, date formats are \emph{Day-Months (3 first letters)-Year (2 last digits)}, e.g., `31-DEC-13'. We reformatted it as \emph{Year (4 digits)-Month (2 digits)-Day}, e.g., `2013-12-31'. Such conversion makes it more suitable to define filtering intervals on dates, using operator such as greater or less than (e.g. date higher than  `2013-06-01' and less than `2013-12-31' to get CANs from the last six months of 2013). Country related fields were converted from the ISO code (2 letters, e.g., `FR') to their full names (e.g., `France'). All award value fields were converted from floats to integers. Finally, award notice IDs were converted to hyperlinks linking to the CAN page on the official TED Web site \cite{TED}. 

Finally, we converted the CSV data to the Apache Parquet format \cite{Parquet,36632}. Apache Parquet is a columnar data storage format designed to support very efficient compression and encoding schemes. Besides compression, Parquet also allows data to be queried from the files using SQL like syntax, a feature which we use in the browser. Data types were associated to CAN fields in order to perform filtering and SQL queries, which we detail in Table \ref{CANdata}. All fields related to numbers were associated an \emph{Integer} data type. Fields involving strings were associated a \emph{String} data type, or \emph{factor} when the number of values was less than 300 hundreds. The use of the factor type allows to present users with a list of choices in the TED browser filtering tool. After Parquet conversion, the dataset size was reduced to 315MB. 

We make available the code for data preparation and Parquet conversion as an IPython notebook \cite{IPythonOpenTED}.

\section{OpenTED browser}
\label{TEDBrowser}

The OpenTED browser is a Web application that provides a user-friendly access to the CANs. Its main features are a filtering tool for extracting subsets of CANs, and a visualisation tool that displays the relationships between contracting authorities and contractors by means of a Sankey diagram. The OpenTED browser is made available online at \cite{OpenTEDBrowser}.

\subsection{Filtering tool}
\label{filtering}

\begin{figure}[!t]
\centering
\includegraphics[width=1\textwidth]{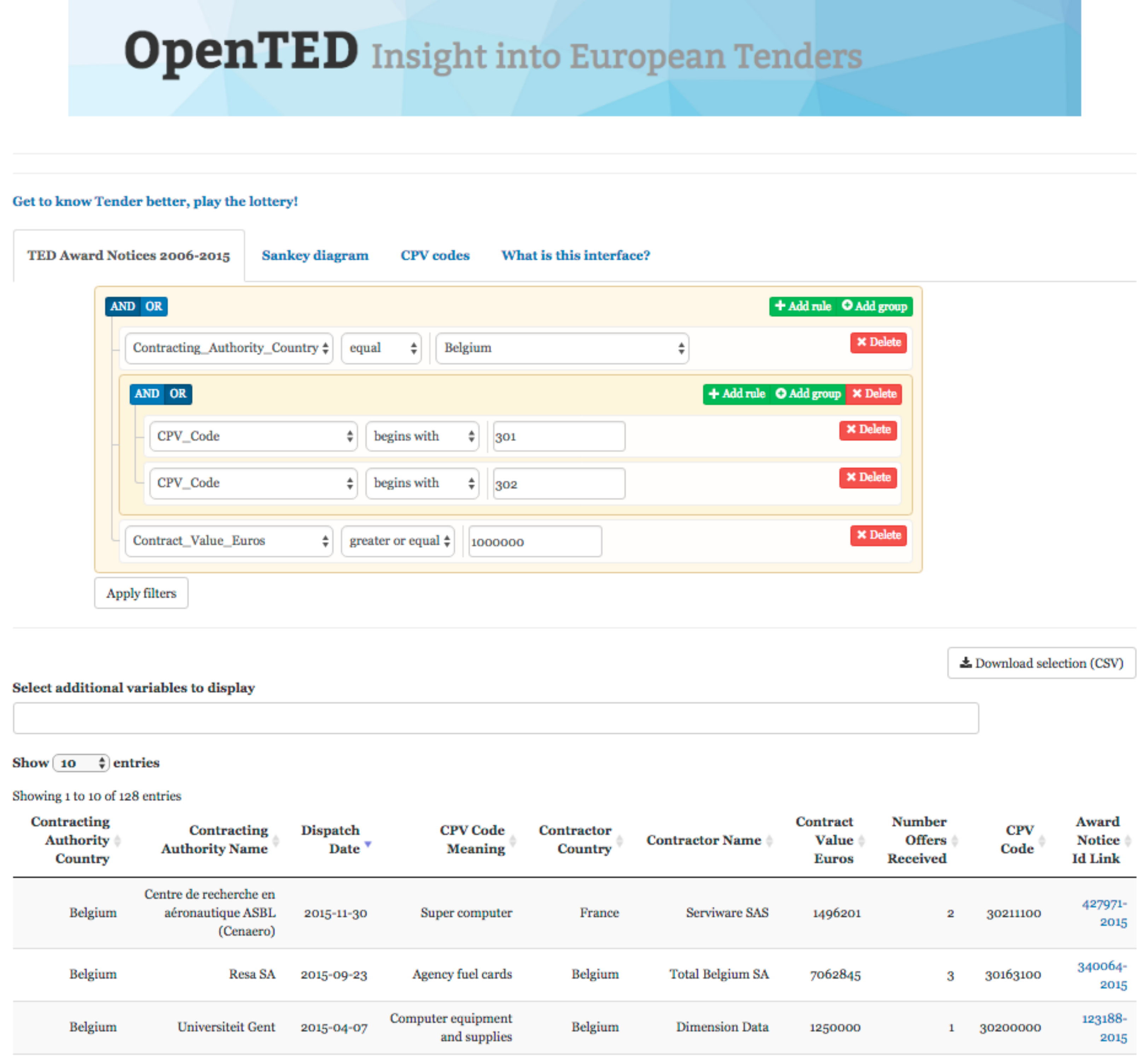}
\caption{OpenTED browser: Filtering tool. The user can filter CANs by setting conditions on CAN field values. A table displays the corresponding subset of CANs.}
\label{UI1}
\end{figure}

The filtering tool allows to filter CANs by setting conditions on the content of any of the CAN fields listed in Tables \ref{CANdata2} and \ref{CANdata}. A snapshot of the tool is given in Fig. \ref{UI1}. All 48 fields can be filtered. The filtering operators provided for a field depends on the data type of the field. A summary of available filtering operators for a given data type is provided in Table \ref{filterOperators}. 

\begin{table}[ht]
\centering
\caption{Available operators for the different data types.}
      \scriptsize
      \begin{tabular}{|l|l|}
        \hline
         Data type   & Available operators \\ \hline
        String & equal,not\_equal,less, less\_or\_equal, greater,greater\_or\_equal,\\
        & between, in, not\_in,begins\_with, ends\_with, is\_null, is\_not\_null \\
        Factor & equal,not\_equal,is\_null, is\_not\_null \\
        Integer & equal,not\_equal,less, less\_or\_equal, greater,greater\_or\_equal,\\
        & between,in, not\_in,is\_null, is\_not\_null  \\
        \hline
      \end{tabular}
\label{filterOperators}
\end{table}

Conditions can be combined by logical conjunction and disjunction operators, and may also be nested thanks to the grouping option. An example of filtering is given in Fig. \ref{UI1}. The filter retrieves CANs for which: (i) the contracting authority country is `Belgium', (ii) the CPV code either begins with `301' (Office machinery, equipment and supplies except computers, printers and furniture) or `302' (Computer equipment and supplies) and (iii) the contract value in euros is more than one million. 

Applying the filter returns the set of CANs matching the conditions. In the example above, 128 CANs are returned, and displayed as a table below the filtering widget. The table is interactive, and the user can reorder columns by increasing or decreasing order of values. Award notice IDs are hyperlinks that connect to the page of the CAN on the official TED Web site \cite{TED}. Finally, the set of filtered CANs can be downloaded as a CSV using the `Download selection' button. 

\subsection{Sankey diagram}

Sankey diagrams visualise the magnitude of flow between nodes in a network. They provide a useful visualisation tool for contract award notices, where contracting authorities and contractors can be seen as nodes of a network, and the contract values as `flows'. 

\begin{figure}[!h]
\centering
\includegraphics[width=1\textwidth]{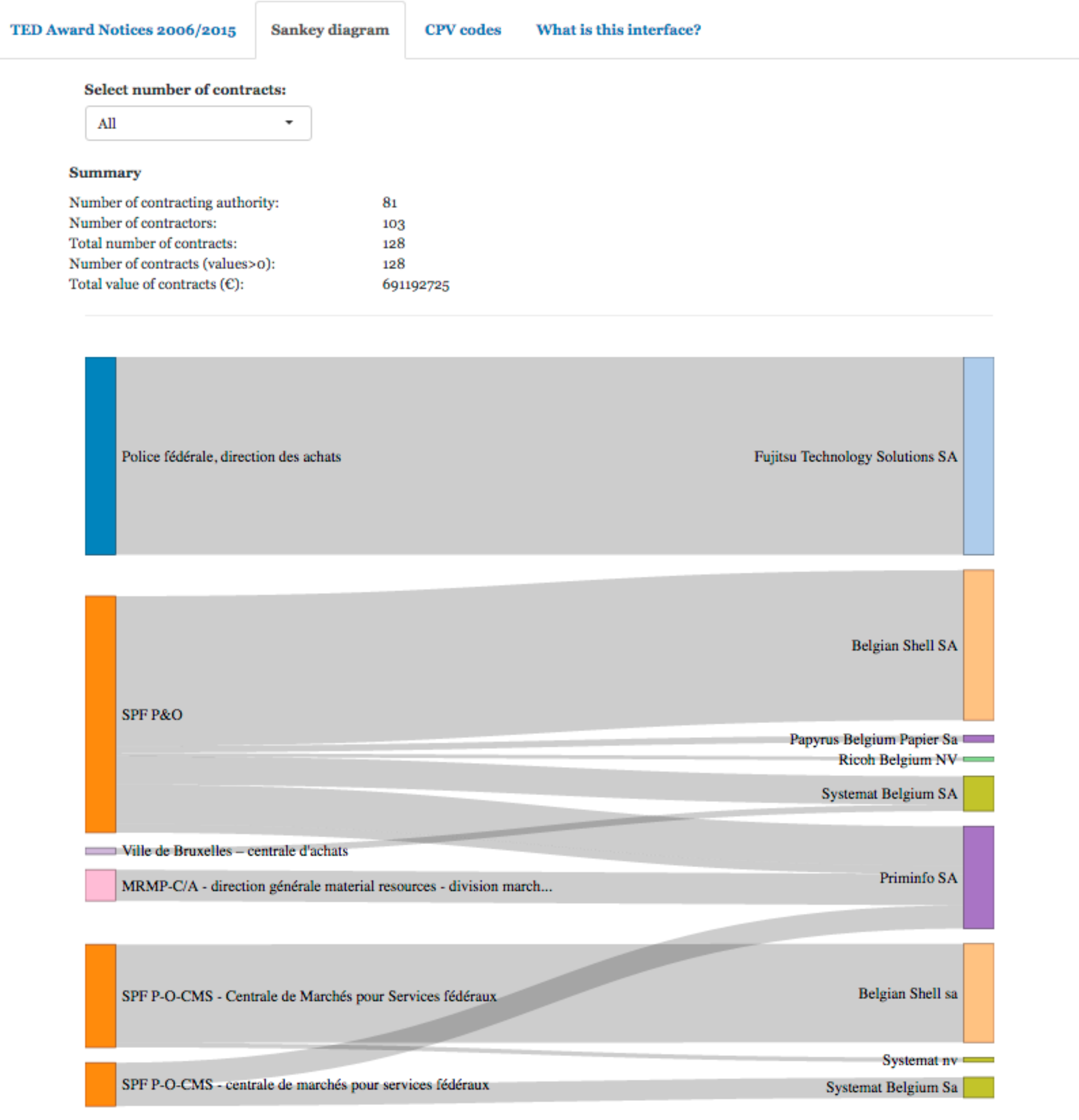}
\caption{OpenTED browser: Sankey diagram for the visualisation of CANs. Contracting authorities and contractors are represented on the left and right sides of the network, respectively. The thickness of flows is proportional to the sum of award values between two parties.}
\label{UI2}
\end{figure}

The set of contracting authorities are represented on the left side of the network, and the set of contractors on the right side. The thickness of the flow is proportional to the sum of contract values between contracting authorities and contractors. 

Fig. \ref{UI2} gives a snapshot of the Sankey diagram obtained for the subset of CANs matching the conditions given in Section \ref{filtering}, and illustrates that the diagram gives a clear overview of the relationships existing between the different parties. A few statistics at the top of the page summarise the number of contracting authorities, contractors, contracts and the sum of contract values in the Sankey diagram.

Tooltips and hyperlinks are tied to the diagram edges, giving the amount of the contract value, and linking to the page of the contract award notice on the official TED Web site, respectively. 

\subsection{Common Procurement Vocabulary}

The subjects of procurement contracts are encoded using Common Procurement Vocabulary (CPV) \cite{CPV}. CPV is based on a tree structure comprising codes of up to 9 digits (an 8 digit code plus a check digit) associated with a wording that describes the type of supplies, works or services forming the subject of the contract. The first two digits identify the general division, and subsequent digits refine the division into more specific categories. For example, all codes starting with \emph{30} are from the general division of \emph{Office and computing machinery, equipment and supplies except furniture and software packages}, while the code \emph{3012453} refer more specifically to \emph{Scanner transparency adapters}.

\begin{figure}[!h]
\centering
\includegraphics[width=1\textwidth]{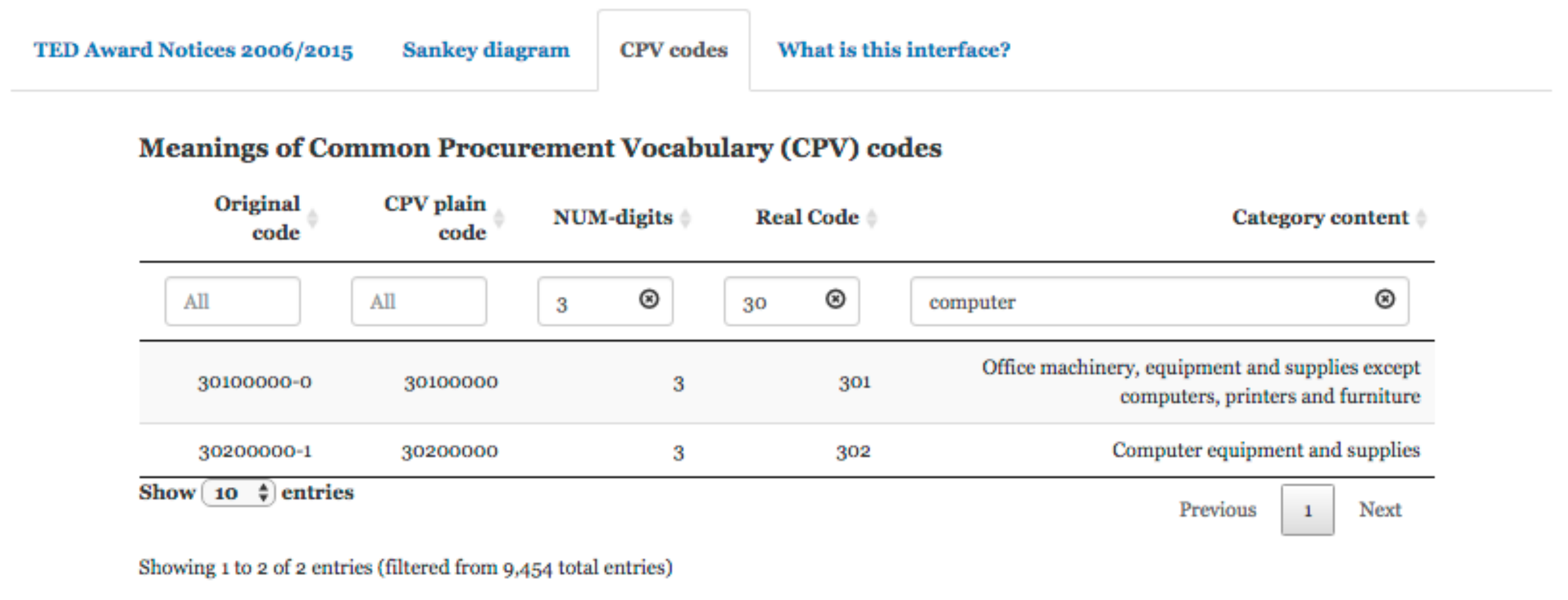}
\caption{OpenTED browser: Meanings of Common Procurement Vocabulary (CPV) codes.}
\label{CPV}
\end{figure}

The meaning of CPV codes can be found from the EU Publications office \cite{CPV}. In order to facilitate the use of CPV codes in the browser, we added a table with all current CPV codes (9454 in total), that can be searched and filtered. The table in available in the \emph{CPV codes} tab, a snapshot of which is given in Fig. \ref{CPV}. The user may fix the number of digits in the results in order to restrict the search to more general categories, and may sort the columns by increasing or decreasing order of values. It should be noted that we only include the current nomenclature (valid since 2008), and that the integration of the previous nomenclature (before 2008) is part of future work. 

\subsection{Implementation}

The Web application was developed using R Shiny \cite{shiny}, and is made Open Source at \cite{IPythonOpenTED}. It is worth mentioning that thanks to the expressiveness of R Shiny, the application is rather compact, i.e. less than 500 lines of code in total, making it easily reusable and adaptable. We furthermore provide a Docker container for facilitating its deployment on a different server, or run it on a local machine for faster interaction with the browser \cite{OpenTEDBrowser}.

\section{Lottery game: Who wants to be a supplier?}

\begin{figure}[!h]
\centering
\includegraphics[width=0.95\textwidth]{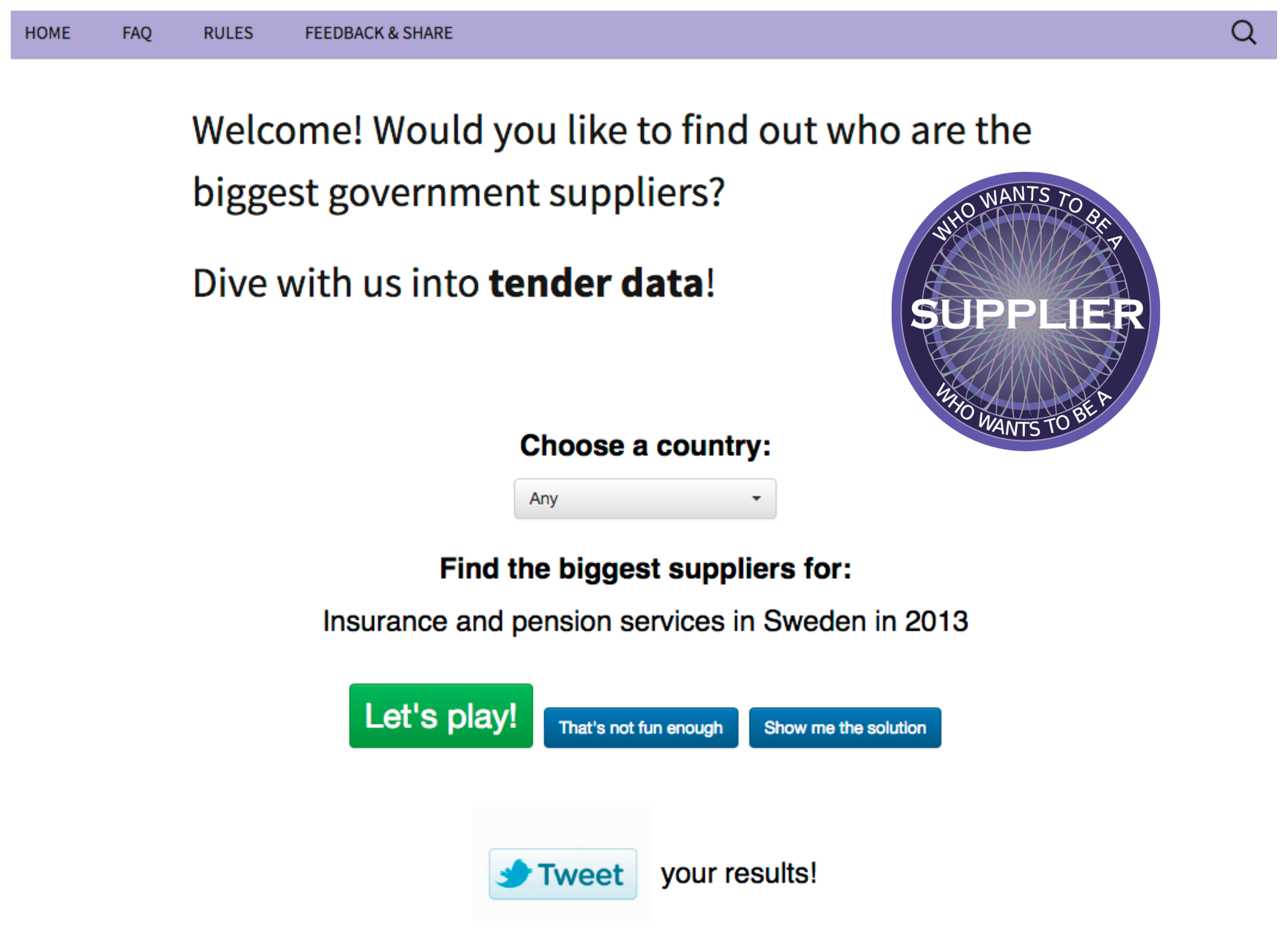}
\caption{Lottery game: Public spending is fun! Who wants to be a supplier?}
\label{lottery}
\end{figure}

The lottery game is a third-party Web site aimed at promoting the OpenTED browser \cite{tenderexposed}. The Web site gives the player a quest on tender data for finding the biggest suppliers of some goods or services in a given country. Fig. \ref{lottery} gives a snapshot of the page, where the quest is to find suppliers for `Insurance and pension in Sweden in 2013'. 

If the player accepts (``Let's play'' button), she is redirected to the OpenTED browser where the goal is to find the subset of CAN corresponding to the quest. She may also get the solution by selecting `Show me the solution', which will also redirect to the OpenTED browser, but the filtering tool will be prefilled with the set of conditions answering the quest. If the user does not like the quest, she can ask for another one by selecting the `Not fun enough' button.

\section{Conclusion and perspectives}

The OpenTED browser provides an intuitive online gateway for filtering contract award notices (CANs) of the European procurement system, and for getting insights into the business relationships existing between contracting authorities and entities. We believe that the tool, thanks to its simplicity and ease of use, can be of significant interest for a number of users. 

Ongoing improvements concern the correction of inconsistencies in the data, related to CPV codes and to the naming of contracting authorities and entities. The nomenclature for CPV codes changed in 2008, which requires to adapt queries to two nomenclatures when searching for CANs before and after 2008. We plan to address this issue shortly. The second type of inconsistency relates to variations in the naming of bidders and buyers. These may be caused by typos, but are also due to different ways of naming an entity (e.g., Siemens, Siemens A.G., Siemmens A.G., and so forth). While efforts to provide unique identifiers instead of names are being promoted, name inconsistencies currently remain an important challenge to properly group CANs according to their contracting authorities and entities. 

In a larger perspective, a wide range of avenues exist for improving the browsing of TED notices, and for providing better insights into tender data. To name a few, possible extensions concern the integration of contract notices, also recently made available by the Publications Office as curated Open Data, and make use of advanced analysis techniques such as clustering, graph analysis, or outlier detection, to investigate questions such as what makes a bidder successful, what are the tendering patterns among EU countries, or to identify indicators of fraudulent behaviours. 

\subsubsection*{Acknowledgments.} The authors acknowledge the support of ``BruFence: Scalable machine learning for automating defense system'' (RBC/14 PFS-ICT 5), a project funded by the Institute for the Encouragement of Scientific Research and Innovation of Brussels (INNOVIRIS, Brussels Region, Belgium), Journalismfund.eu for organising the DataHarvest/European Investigative Journalism Conference, the OpenTED working group (http://ted.openspending.org), the European Union Publications Office, http://ted.europa.eu, 1998--2016, and the European Commision, Directorate-General for Internal Market, Industry, Entrepreneurship and SMEs (DG-GROW) for providing the data, and J\'achym Hercher, Policy Officer at DG-GROW, for providing feedback and improvements on this article. 

\bibliographystyle{splncs03} 
\bibliography{SoGood}      

\section*{Appendix}
 
\noindent\begin{minipage}{\linewidth}
\centering
\captionof{table}{Fields present in contract award notices CSV files}
      \scriptsize
      \begin{tabular}{|l|l|l|}
        \hline
         Field   & Description & Type\\ \hline
         \rowcolor{Gray}
        Notice metadata & & \\
        \textbf{ID\_NOTICE\_CAN} & Unique identifier of the contract award notice & String\\
        YEAR & Year of publication of the notice & Integer \\
        ID\_TYPE & Standard form number  & Factor \\ 
        \textbf{DT\_DISPATCH} & The date when the buyer dispatched (sent) the notice & String\\
        & for publication to TED & \\ 
        XSD\_VERSION &Version of the XML schema definition [ADDED] & Factor\\ 
        CANCELLED &1 = this notice was later cancelled [ADDED] & Factor\\ 
         \rowcolor{Gray}
        Contracting authority or &&\\
         \rowcolor{Gray}
        entity identification && \\
        \textbf{CAE\_NAME} & Official name & String \\ 
        CAE\_NATIONALID &``National ID" e.g. VAT number for utilities  & String \\ 
         CAE\_ADDRESS  &Postal address & String\\ 
         CAE\_TOWN & Town & String  \\
        CAE\_POSTAL\_CODE & Postal code  & String \\ 
       \textbf{ISO\_COUNTRY\_CODE} & Country & Factor \\
          \rowcolor{Gray}
        Winning bidder identification & & \\
         \textbf{WIN\_NAME} & Official name & String \\
          WIN\_ADDRESS & Postal address& String\\
         WIN\_TOWN  &Town& String\\
          WIN\_POSTAL\_CODE  & Postal code& String\\
          \textbf{WIN\_COUNTRY\_CODE}& Country& Factor \\
         \rowcolor{Gray}
        Various CAN level variables&  & \\
         CAE\_TYPE & Type of contracting authority & Factor\\
         MAIN\_ACTIVITY & The classification corresponds to COFOG divisions & String \\
          B\_ON\_BEHALF  &This indicates either a central purchasing body or & Factor\\
          &  several buyers buying together & \\
          TYPE\_OF\_CONTRACT & Type of contract & Factor \\
         TAL\_LOCATION\_NUTS & The Nomenclature of Territorial Units for & String\\
         &Statistics (NUTS) code placement & \\
          B\_FRA\_AGREEMENT & The notice involves the establishment of a & Factor\\
          & framework agreement & \\
          B\_DYN\_PURCH\_SYST&  The notice involves contract(s) based on a dynamic & Factor\\
&purchasing system &\\
          \textbf{CPV} &  The main Common Procurement Vocabulary code of & String\\
&the main object of the contract &\\
          ADDITIONAL\_CPV1-4 & The first four CPV listed in the notice. & String\\
          B\_GPA & The contract is covered by the Government & Factor\\
&Procurement Agreement &\\
          \textbf{VALUE\_EURO} & CAN value, in EUR, without VAT. If the value was & Integer\\
&not present, the lowest bid is included &\\
          VALUE\_EURO\_FIN\_1 & CAN value, in EUR, without VAT. If the value was& Integer\\
&not present, second estimate&\\
          VALUE\_EURO\_FIN\_2 & CAN value, in EUR, without VAT. If the value was& Integer\\
&not present, third estimate &\\
        TOP\_TYPE  & Type of procedure & Factor\\
         CRIT\_CODE &  Award criteria & Factor\\
          CRIT\_CRITERIA&  Information on award criteria. & String\\
          CRIT\_WEIGHTS& Information on award criteria weighing & String \\
          B\_ELECTRONIC\_AUCTION & An electronic auction has been used & Factor \\
          NUMBER\_AWARDS & The number of CAs for a given CAN. [ADDED] & Integer\\
            \rowcolor{Gray}
        Various CA level variables& & \\
          ID\_AWARD& Unique contract award identifier & String \\
        CONTRACT\_NUMBER& Contract No& String \\
        LOT\_NUMBER &   An identifier of a lot & String\\
        TITLE &  Title & String\\
        \textbf{NUMBER\_OFFERS}& Number of offers received & Integer\\
        NUMBER\_OFFERS\_ELECTR&  Number of offers received by electronic means & Integer\\
       AWARD\_EST\_VALUE\_EURO & Estimated CA value, in EUR, without VAT & Integer\\
        AWARD\_VALUE\_EURO & Total final CA value, in EUR, without VAT. If the& Integer\\
&value was not present, the lowest bid is included& \\
        AWARD\_VALUE\_EURO\_FIN\_1 &  CA value, in EUR, without VAT. If a value field is& Integer\\
&missing, second estimate&\\
       B\_SUBCONTRACTED  &  The contract is likely to be subcontracted & Factor\\
       B\_EU\_FUNDS  &  The contract is related to a project and / or& Factor\\
&programme financed by European Union funds& \\
        DT\_AWARD &  Date of contract award& String\\
        \hline
      \end{tabular}
\label{CANdata}
\end{minipage}

\end{document}